# Interlayer Sliding-Induced Intralayer Ferroelectric Switching in Bilayer Group-IV Monochalcogenides

Bo Xu[1], Junkai Deng[1, *], Xiangdong Ding[1], Jun Sun[1] and Jefferson Zhe Liu[2, *]

[1]*State Key Laboratory for Mechanical Behavior of Materials, Xi'an Jiaotong University, Xi'an 710049, China*

[2]*Department of Mechanical Engineering, The University of Melbourne, Parkville, VIC 3010, Australia*

Email: junkai.deng@mail.xjtu.edu.cn; zhe.liu@unimelb.edu.au

## Abstract

Two-dimensional materials with ferroelectric properties break the size effect of conventional ferroelectric materials and unlock unprecedented potentials of ferroelectric-related application at small length scales. In this work, using density functional theory (DFT) calculations, we discover a tribo-ferroelectricity behavior in a group of bilayer group-IV monochalcogenides (MX, with M = Ge, Sn and X = S, Se). Upon interlayer sliding over an in-plane unit cell length, the top layer exhibits a reversible intralayer ferroelectric switching, leading to a reversible transition between the ferroelectric (electric polarization of 40 $\mu C/cm^2$) and antiferroelectric states in the bilayer MXs. Our results show that the interlayer van der Waals interaction, which is usually considered to be weak, can actually generate an in-plane lattice distortion and thus cause the breaking/forming of intralayer covalent bonds in the top layer, leading to the observed tribo-ferroelectricity phenomenon. This unique property has several advantages for energy harvesting over existing piezoelectric and triboelectric nanogenerators. The interlayer sliding-induced polarization change is as high as 40 $\mu C/cm^2$, which can generate an open-circuit voltage two orders of magnitude higher than that of $MoS_2$-based nanogenerators. The polarization change occurs over a time period for interlayer sliding over a unit-cell length, leading to an ultrahigh polarization changing rate and thus an ultrahigh short-circuit current. The theoretical prediction of power output for the tribo-ferroelectric bilayer MXs at a moderate sliding speed 1 m/s is four orders of magnitude higher than the $MoS_2$ nanogenerator, indicating great potentials in energy harvesting applications.

**Introduction**

Ferroelectric (FE) materials possessing stable and switchable spontaneous electric polarization are essential elements in many technology applications, including memories, field-effect transistors, solar cells, sensors, and actuators. Over the past decade, advances in two-dimensional (2D) functional materials have stimulated broad research interest in 2D ferroelectrics[1]. Compared with traditional FE materials, 2D ferroelectrics are more easily integrated as functional components in miniaturized electronic devices owing to their intrinsic nanoscale size and ferroelectricity without the limitation of the size effect[2]. Following extensive theoretical predictions of 2D FE materials in the past few years, ferroelectricity has been experimentally observed in some 2D materials with thicknesses as low as several unit cells and even one unit cell[2–6]. Similar to conventional ferroelectrics, the intrinsic intralayer polarization of 2D materials usually originates from an asymmetric crystal structure caused by atom displacement and/or asymmetric charge redistribution. For example, a distorted SnTe monolayer was experimentally detected, giving rise to a robust in-plane intralayer polarization[2]. The displacement of Mo atoms breaks the inversion symmetry of 1T-$MoS_2$ and results in an out-of-plane intralayer polarization[7]. A similar atomic displacement and lattice distortion-induced polarization has also been found in $CuInP_2S_6$[3], $In_2Se_3$[8], and elemental group-V monolayer materials[9]. The intralayer electric polarization of 2D FE materials can be as high as 48 $\mu C/cm^2$, which is comparable with that of conventional FE materials[10].

Moreover, 2D materials offer new artificial designation approaches, e.g., interlayer sliding, to induce non-intrinsic ferroelectricity owing to their unique layered crystal structures. The relatively weak interlayer van der Waals (vdW) interaction allows 2D materials to stack freely without being limited by the directional atomic bonds or lattice mismatch and endows these materials with diverse stacking-dependent properties[11–14]. Tuning stacking order can break the centrosymmetric nature of the crystal and induce interlayer charge transfer in bilayers or few-layer 2D materials, such as $WTe_2$[5], $VS_2$ bilayers[15], and BN bilayers[16], which then leads to an out-of-plane interlayer electric polarization. Interlayer sliding is predicted to change the interlayer potential and switch the polarization from upward to downward[16–18]. This is an interesting phenomenon since it provides a new way to control the FE polarization in addition to the conventional electric field. Such unique properties could stimulate novel concepts for various applications, such as triboelectric nanogenerators (NGs) for energy harvesting. However, the interlayer charge transfer is weak[19], and the obtained FE polarization is less than 0.5 $\mu C/cm^2$, i.e., two orders of magnitude smaller than that of conventional FE $BaTiO_3$ (30 $\mu C/cm^2$). The dilemma is the following: while, on one side, the weak interlayer interaction gives rise to interlayer sliding-controlled ferroelectrics, on the other side, it severely limits the polarization magnitude and thus the applications.

In light of the large intralayer FE polarization of some 2D materials[10,20–22], a natural question arises: is it possible to use interlayer sliding to switch the intralayer polarization and thus solve the dilemma? Intuitively, this sounds very unlikely since the interlayer vdW interaction is very weak, whereas intralayer atom displacement is usually constrained by strong chemical bonds. In this work, using density functional theory (DFT) calculations, we prove this hypothesis for bilayer group-IV monochalcogenides, namely MX (M = Ge, Sn; X = S, Se), thus revealing a huge potential for 2D ferroelectrics. MXs have a distorted puckered structure that generates a considerably large intralayer FE polarization in all predicted 2D ferroelectrics[1,23]. Our results show that a mechanical interlayer sliding can trigger a direction change of the electric polarization of the top layer, leading to a reversible transition between the FE and antiferroelectric (AFE) states of bilayer MXs. Our in-depth analysis shows that, under different stacking orders of bilayer MXs, the out-of-plane interlayer vdW interaction can induce an in-plane lattice distortion and thus promote intralayer covalent bond breaking/forming in the top layer. This new tribo-FE/AFE phenomenon shows that the robust intralayer FE polarization in 2D materials can be switched not only by external fields, as is traditionally done[24–27], but also by this unique interlayer sliding mechanism. Under a sliding speed of 1 m/s, the large polarization switching occurring within the nanoscale atomic lattice leads to an alternating displacement current with ultrahigh frequency. For example, a bilayer MX flake with a length of 10 nm can in principle generate a short-circuit current of 35 nA and a maximum power output of 107 nW, which is four orders of magnitude higher than the state-of-the-art $MoS_2$-based NG[28]. The superior electrical performance renders bilayer MXs promising candidates for potential application as NGs and energy harvesting devices. Through the recently obtained advances in 2D material tribology, many techniques have been developed to precisely manipulate interlayer sliding[29–33]. Therefore, it is of great relevance to implement these findings in real-world applications.

## Results and Discussion

### Stacking order of bilayer MX and FE/AFE state

Monolayer group-IV monochalcogenides, namely MX (M = Ge, Sn; X = S, Se), have a hinge-like crystal structure similar to that of black phosphorene. Due to the broken centrosymmetry in the monolayer, a considerable spontaneous electric polarization, ranging from 18 to 48 μC/cm², exists along the *x* (armchair) direction (Fig. 1a) and is retained at room temperature[22,26,34,35]. For bilayer MXs stacked together by two monolayers, as shown in Fig. 1b, there are four high-symmetry stacking orders. The AA stacking order refers to the top layer being precisely matched with the bottom layer in the *x-y* plane. Compared with the AA stacking, the AB stacking corresponds to the top layer being shifted by a half lattice constant along the *x*-direction, i.e., by 0.5*a* (where *a* is the lattice constant in the armchair direction). Shifting the top layer by a half lattice constant 0.5*b* along the *y*-direction (where *b* is the lattice constant in the zigzag direction) yields the AC stacking. Finally, moving the top layer by a half lattice constant along both the *x*- and *y*-direction results in the AD stacking. As each monolayer exhibits intrinsic FE polarization, bilayer MX can be constructed with either parallel or anti-parallel polarization coupling between the two layers for a given stacking order. Consequently, a specific stacking order exhibits either an FE state or an AFE state (Fig. 1c).

The energetic orders of these bilayer MXs for these four staking orders and corresponding two electric polarization states were investigated via DFT calculations. In these DFT calculations, the relative position of the two layers was fixed, while the lattice constants and interlayer spacing were allowed to relax. Supplementary Table 1 summarizes the results. For bilayer MXs under the FE state, the AC stacking order, here denoted by $AC_{FE}$, is the most stable. The general energetic order trend is as follows: $AC_{FE} < AB_{FE} < AA_{FE} < AD_{FE}$. By contrast, for the AFE state, bilayer MXs with the $AB_{AFE}$ stacking order are the most stable, and $AC_{AFE}$ is the second most stable stacking order structure. GeSe, SnS, and SnSe bilayers have $AA_{AFE} < AD_{AFE}$, which is similar to the FE state. However, the opposite is true for GeS bilayers.

Fig. 1c shows the energy difference between the FE and AFE states for each stacking order. For all these bilayer MXs, the AA and AC stacking orders prefer the FE state, but the AB and AD stacking orders prefer the AFE state. Although the bulk allotrope of these four MX materials has an AFE configuration, both the FE and AFE states in 2D nanoscale materials have been successfully fabricated and characterized in experiments[25,27,36,37]. Based on these results, we speculate that the change of stacking order via mechanical interlayer sliding (in tribology) could lead to the intralayer polarization switching of top layer, thus, for bilayer MXs, corresponding to a phase transition between the FE and AFE states. Such a phase transition and resultant polarization change/switch has not yet been reported.

**Interlayer sliding-induced reversible FE–AFE transition and polarization switching**

Using DFT calculations, the total energies and the polarization variation (in the *x*-direction) of bilayer MX were examined through a progressive mechanical sliding of the top layer with respect to the bottom layer. The $AC_{FE}$ state was taken as the starting point. Note that, owing to the significant advances obtained in 2D material tribology in the past few years, precise control of mechanical interlayer sliding in various 2D materials has been achieved[29–31].

Bilayer SnSe is here taken as an example to show the polarization variation along two sliding pathways that connect two high symmetry stacking orders: from AC to AD in Fig. 2a and from AC to AB in Fig. 2b. Note that bilayer SnSe exhibits a stable FE state for the AC stacking and a stable AFE state for both AD and AB stackings (Fig. 1c). As the top layer moves toward the AD stacking along the $x$-direction (Fig. 1b), Fig. 2a shows a gradual reduction of $P_s$ from 40 to ~30 μC/cm$^2$ followed by an abrupt drop to nearly zero at a sliding distance of 0.128$a$. After careful inspection of the crystal structure in Supplementary Figure 1a, it was found that the top layer had switched the polarization direction via breaking and forming of Sn–Se covalent bonds. The anti-parallel polarization of the two layers forms an AFE state. This AFE state remains unchanged until the top layer continues to slide into the AD$_{AFE}$ state. In the reverse sliding direction (starting from the AD$_{AFE}$ state), bilayer SnSe remains in the AFE state until the relative sliding distance reaches 0.021$a$. At this critical point, a sharp increase of $P_s$ from zero to ~40 μC/cm$^2$ is observed, indicating a phase transition back to the FE state. Indeed, the analysis of the crystal structures shows that the polarization of the top layer switches back to the initial state during this backward sliding process (Supplementary Figure 1a). Therefore, the observed phase transition between AC$_{FE}$ and AD$_{AFE}$ is reversible and occurs spontaneously. The different critical phase transition points in these two opposite sliding directions result in a hysteresis loop, as shown in Fig. 2a. Similarly, Fig. 2b shows the pathway from AC to AB. $P_s$ decreases slightly to 35 μC/cm$^2$ and then drops abruptly to nearly zero at a sliding distance of 0.115a in the $x$-direction. Again, via inspection of the atomic structures of Supplementary Figure 1b, a crystal structure change of the top layer similar to that of the AC–AD case is observed, i.e., breaking and subsequent forming of Sn–Se covalent bonds. A backward mechanical sliding results in a sudden increase of $P_s$ to 40 μC/cm$^2$ at 0.06$a$, indicating a spontaneous reverse phase transition to the FE state.

Moreover, the relative total energy change versus interlayer sliding was investigated. The bottom plot of Figure 2a shows the case of the corresponding path from AC to AD. The total energy of AC$_{FE}$ is here taken as reference. Starting from AC$_{FE}$, the relative energy shows a parabolic-like increase until a sliding distance of 0.128$a$, at which point the energy suddenly drops, corresponding to the FE-to-AFE phase transition. The energy then increases following another smooth curve till the AD stacking is reached. It should be noted that AD$_{AFE}$ is at the top of the energy curve and is thus mechanically unstable. Once the mechanical sliding constraints applied to the top layer are released, the top layer slides spontaneously in the backward direction. In the backward sliding process, the bilayer SnSe energy follows the second smooth energy curve till 0.021$a$, at which point the energy drops, corresponding to the AFE-to-FE phase transition in the curve of polarization. These two smooth energy curves clearly correspond to the FE (black) and AFE (red) states, respectively. They have a crossing point at ~0.08$a$, which separates the stable FE and AFE regions. Note that the two observed spontaneous phase transitions do not take place at this crossing point. This can be attributed to the presence of an energy barrier that will be discussed later. The bottom plot of Figure 2b shows the variation of the relative total energy along the AC–AB sliding pathway. In contrast to AD$_{AFE}$ in Fig. 2a, AB$_{AFE}$ is a metastable structure with an energy barrier of 18 meV/atom that separates it from the ground state AC$_{FE}$. Upon releasing the sliding constraints, bilayer SnSe remains in the AB$_{AFE}$ stacking order. This phenomenon could be used in information storage applications.

Furthermore, comprehensive DFT calculations were conducted to obtain the variation of the electric polarization and total energy as a function of the interlayer sliding in the whole $x$-$y$ plane. Figure 2c, d shows the polarization-sliding map for SnS and SnSe, respectively. Supplementary Figure 2 shows the total energy landscape. Note that these two maps are obtained by progressively sliding the top layer along various sliding pathways (starting from the AC$_{FE}$ state) in DFT calculations. For SnS and SnSe, neither FE nor AFE has a complete energy landscape across the whole $x$-$y$ sliding plane due to the spontaneous phase transition. Therefore, the FE and AFE energy landscapes were combined together (Supplementary Figure 2), and the phase boundary is depicted with a dotted line in Fig. 2c, d. At the boundary, the FE and AFE states have equal total energy values. The middle region of the $x$-$y$ sliding plane represents the stable AFE state. AC$_{FE}$ and AB$_{AFE}$ are stable states for SnS and SnSe, whereas AA$_{FE}$ and AD$_{AFE}$ are unstable states. Across the phase boundary, there is a significant $P_s$ drop from 40–60 $\mu C/cm^2$ to zero (Fig. 2c, d).

Similarly, Fig. 2e, f shows the polarization-sliding map of GeS and GeSe, and Supplementary Figure 3 illustrates the corresponding energy landscapes. Through DFT calculations, one energy landscape is obtained for the FE state and another one for the AFE states across the whole $x$-$y$ sliding plane (Supplementary Figure 3a, b, d, e). The energy differences between the states are shown in Supplementary Figure 3c, f for GeS and GeSe, respectively. The intersection curves of the two energy landscapes are projected onto the $x$-$y$ plane and indicated by dashed lines. They represent the theoretical phase boundaries that separate the stable FE and AFE regions. Combining the phase diagram (Supplementary Figure 3) and polarization contour map (Supplementary Figure 4), the theoretical polarization phase diagrams for GeS and GeSe can be obtained, as shown in Fig. 2e, f, respectively. A sharp change of $P_s$ takes place across the boundary. Due to the energy barrier, the FE-to-AFE phase transitions cannot spontaneously take place during mechanical sliding, but can occur at a specific finite temperature.

The nudged elastic band (NEB) method was adopted in the DFT calculations to determine the energy barrier (Supplementary Figure 5). In the vicinity of the phase boundary, the energy barriers of bilayer GeS and GeSe are about 78 and 33 meV/atom, respectively. They are comparable with the thermal excitation energy at ambient conditions (~26 meV/atom at 300 K). Thus, GeSe could exhibit the phase transition near room temperature, whereas GeS might need an elevated temperature (around 900 K) for the phase transition to occur. From a thermodynamics perspective, mechanical sliding across the boundaries would lead to a reversible FE/AFE phase transition. In contrast, SnS and SnSe have an almost vanishing energy barrier near their phase boundary, which is consistent with the observed spontaneous phase transition.

Our DFT calculations clearly indicate that the FE/AFE phase transition is feasible upon mechanical interlayer sliding for these bilayer MXs. Such a tribo-ferroelectricity is distinctive from traditional ferroelectricity and ferroelectricity reported in 2D ferroelectrics, where the electrical polarization is usually manipulated by an external electric field[2,26,27]. Indeed, the ferroelectricity observed in this work originates from the layered crystal structures of vdW materials and the corresponding easy mechanical interlayer sliding. Additionally, the electric polarization can be changed from 0 to about 40–90 $\mu C/cm^2$, which is higher than the polarization of the more commonly used $BaTiO_3$, which exhibits a $P_s$ of ~30 $\mu C/cm^2$ in its tetragonal phase[38,39], and of the interlayer ferroelectricity (less than 0.1 $\mu C/cm^2$) in bilayer $WTe_2$, whose polarization originates from the asymmetric stacking-induced charge transfer between layers[5,17,19].

**Physical origins of the observed tribo-FE phenomena**

The FE/AFE phase transition of the discovered tribo-FE phenomenon involves covalent bonds breaking and forming within the MX layer (Supplementary Figure 1). This is surprising as conventional knowledge suggests that interlayer vdW interactions are much weaker than intralayer chemical bonds. In the past, to change the intralayer crystal structure and thus achieve polarization switching, either an external stress or an electric field had to be directly applied to the given layer[10,24,26]. Note that some recent studies have shown that the interlayer vdW interaction could stabilize stacked crystal structures in some few-layer vdW materials (e.g., the AA stacking of few-layer SnS and the AFE state of $CuInP_2S_6$ group) contrary to their bulk counterparts[27,40,41]. However, the reversible intralayer chemical bonds reconfiguration caused by interlayer interactions has not been reported before. It is thus relevant to investigate the physical origins of this anomalous tribo-FE phenomenon.

A careful inspection of the crystal structures reveals that interlayer sliding leads to structural distortions, including both interlayer vertical distance variation (Supplementary Figure 6) and in-plane lattice strains (Supplementary Figure 7 and Supplementary Table 1). When two fully relaxed monolayers stack together, the interlayer interaction lowers the total energy of the bilayer system, which is directly correlated to the interlayer vertical distance in the sliding process. It should be noted that one stacked layer is slightly different from the previous fully relaxed monolayer because the intralayer bonds and lattice constants are slightly changed by the interlayer interaction. As listed in Supplementary Table 1, the lattice strain in the $x$-direction became as large as 2%. For quantitative analysis, the total energy of bilayer MX was split into three components:

$$E_{bi-MX} = 2E_{MX} + E_\varepsilon + E_{inter},$$

where $E_{MX}$ is the total energy of the fully relaxed monolayer MX, $E_\varepsilon$ is the strain energy of the top and bottom layers, and $E_{inter}$ is the interlayer interaction between the slightly deformed layers. The observed lattice distortion ($E_\varepsilon$) should be attributed to the sliding-induced interlayer interaction changes ($E_{inter}$). Taking SnSe as an example, we calculated the total energy of the rigid bilayer SnSe during its transition from the AC to the AD stacking and compared it with those of the relaxed cases shown in Supplementary Figure 8. Figure 3a shows the $E_{inter}$ and $E_\varepsilon$ results separately. Comparing the relaxed curve (with lattice distortion) with the rigid case (no lattice distortion), it is noticed that the lattice distortion significantly reduces $E_{inter}$ while only slightly increasing $E_\varepsilon$.

To explore the driving force of the tribo-FE/AFE transition, an energy analysis for the FE and AFE phases was conducted for SnSe transitioning from the AC to the AD stacking. Figure 3b, c summarizes the variation of $E_{inter}$ and $E_\varepsilon$. The $E_{inter}$ changes are opposite to the trend of $E_{tot}$ in Fig. 2a, clearly indicating that $E_{inter}$ is not the driving force behind the transition. As for the lattice strain energy $E_\varepsilon$, it can be observed from Fig. 3c that $E_\varepsilon$ increases as the top layer slides away from the $AC_{FE}$ state and then exhibits a sudden drop of 4.2 meV/atom to the AFE state at the phase transition point. In the reverse sliding process, $E_\varepsilon$ gradually increases till the phase transition returns to the FE state, at which point an energy drop of 4.1 meV is observed. The magnitude of the $E_\varepsilon$ change is close to that of $E_{tot}$, indicating that the lattice strain energy relaxation is a key driving force for the tribo-FE phase transition of bilayer SnSe. By further splitting $E_\varepsilon$ into the two different contributions from the top and bottom layers, as shown in Supplementary Figure 9, the top layer strain energy undergoes a minor drop during the phase transition. The observed $E_\varepsilon$ change in Fig. 3c for bilayer SnSe originates from the bottom layer in Supplementary Figure 9a, indicating that the release of strain energy in the bottom layer contributes more to the FE-to-AFE phase transition. In the reverse $AD_{AFE}$-to-$AC_{FE}$ process illustrated in Supplementary Figure 9b, the strain energy of the top and bottom layers decreases by 2.47 and 1.66 meV/atom, respectively, revealing that the release of strain energy for both layers contributes to the reverse phase transition.

The lattice distortion also has a critical role in setting the energy barrier separating the FE from the AFE state. Figure 3d presents the energy barrier results between the FE and AFE states for bilayer SnSe at different sliding distances from $AC_{FE}$ to $AD_{AFE}$. When the sliding distance is far away from the transition point, e.g., $0.05a$, the FE state needs to overcome an energy barrier of 1.2 meV/atom to transform into the AFE state. Upon further sliding, the FE state gradually becomes metastable compared with the AFE state, and the energy barrier keeps decreasing. Upon reaching $0.127a$ (close to the critical point of $0.128a$), the energy barrier is nearly zero. The phase transition thus happens spontaneously, which is consistent with Fig. 2a. In our energy barrier calculations, it was noticed that the transition state has a unit cell with an $a/b$ ratio much closer to one (cubic-like) than those of the initial FE state and final AFE state. To quantitatively investigate the influence of the unit cell shape, the unit cell rectangularity is defined as:

$$\text{Rectangularity} = \frac{a}{b} - 1,$$

where $a$ and $b$ are the lattice parameters in the armchair and zigzag direction, respectively. Figure 3e shows the rectangularity of bilayer SnSe during mechanical sliding. Upon sliding, the rectangularity continuously decreases from about 0.04 to a minimum value of about 0.02 and then shows a sudden jump at the FE-to-AFE phase transition (black line in Fig. 3e). The reverse process is similar, and the rectangularity in this case reaches a minimum value <0.01 at the phase transition (red line in Fig. 3e). These results indicate that a small rectangularity strongly correlates with the disappearance of the energy barrier. Indeed, Fig. 3f shows the quantitative confirmation of such correlation. This figure demonstrates that the energy barrier decreases with decreasing rectangularity. A near-zero energy barrier is achieved when the rectangularity becomes smaller than 0.025. Moreover, the energy barriers of bilayer SnSe (hollow symbols) were compared with those of monolayer SnSe (solid symbols) for the same lattice constants (thus, for the same rectangularity). The comparable energy barrier values clearly indicate that lattice strain is the physical origin behind the energy barrier change rather than the interlayer interaction.

Through the mechanical sliding process, the vdW force leads to a significant lattice strain for bilayer SnS and SnSe, drastically reducing their rectangularity. This gives rise to a vanishing energy barrier and, consequently, to a spontaneous phase transition. On the contrary, as shown in Supplementary Figure 10, the $AC_{FE}$ states of bilayer GeS and GeSe have a large rectangularity, 0.20 and 0.12, respectively. These rectangularity values did not reduce, but instead increased to 0.22 and 0.13, upon sliding close to the phase boundary (Fig. 2). The relatively high rectangularity values are likely the reason for the calculated non-zero energy barrier in the vicinity of the phase boundary in Supplementary Figure 5. For the tribo-FE phase transition to take place, some external stimulus, such as temperature, electric field, or applied stress, is required to assist the interlayer sliding for bilayer GeS and GeSe.

**Performance of the tribo-FE device as NG**

NGs are an emerging technology: they harvest energy from the ambient environment for self-powered micro/nanosystems. It is well established that the energy harvesting performance of NGs depends on the polarization-related displacement current, $J_D = \frac{\partial P_s}{\partial t}$. Polarization change is the key factor for the electrical output[42]. In piezoelectric nanogenerators (PENGs), such as monolayer $MoS_2$, polarization change is obtained by applying an external strain, a phenomenon that is known as piezoelectricity[28,43]. Triboelectric nanogenerators (TENGs) obtain polarized charges through contact electrification (triboelectricity) and convert mechanical energy into electricity via electrostatic induction[44]. Notice that the polarization change (~40 $\mu C/cm^2$) induced by phase transition in tribo-FE is two orders of magnitude higher than that of PENGs (~0.27 $\mu C/cm^2$ for monolayer $MoS_2$), which should lead to a much higher open-circuit voltage (proportional to the polarization change). The time during which the polarization change occurs is only ~0.5 ns under a moderate interlayer sliding speed of 1 m/s. In other words, the rate of polarization change of tribo-FE is ultrahigh and can generate a considerable displacement current in principle. Such an intrinsic superiority motivates us to propose a tribo-FE-based NG and estimate its potential electrical performance. For simplicity, bilayer SnSe flakes with an in-plane size of 10 × 10 nm were employed in this work (see the details in the Supplementary Information).

Taking bilayer SnSe as an example, Fig. 4a shows the schematic illustration of a tribo-FE-based NG in connection with an external load resistor, which is analogous to the previous model[28,44,45]. The tribo-FE/AFE phenomenon in bilayer SnSe results in multiple repeated bursts of a significant electrical polarization change upon continuous interlayer sliding (over distances of several unit cells in Fig. 4a). Fig. 4b illustrates the variation of polarization, source voltage, and short-circuit current of the proposed NG device in three sliding periods. The polarization $P_s$ repeatedly changes between ~40 and 0 $\mu C/cm^2$. Using a well-established theoretical model[42], the source voltage can be estimated for the 10 × 10 nm bilayer MX device in the middle of Fig. 4b, showing an alternating change between ~25 and 0 V, as indicated by the dotted line. For simplicity, the source voltage is approximated as a simple sinusoidal alternating curve (the blue curve). The open-circuit voltage $V_{oc}$ can be obtained from this curve (see details in the Supplementary Information). The considerable output voltage benefits from the significant polarization change (~40 μC/cm²) during the interlayer sliding process. For monolayer MoS₂-based PENGs with the same size[45], the polarization change is only ~0.27 μC/cm², and the voltage is only ~0.24 V. The voltage of the proposed tribo-FE NG (~25 V) is thus two orders of magnitude higher.

For the proposed tribo-FE bilayer device, the polarization change rate depends on the sliding speed $v$, i.e., $\frac{dP}{dt} = \frac{dP}{dx} \cdot \frac{dx}{dt} = \frac{dP}{dx} \cdot v$. The bottom plot of Fig. 4b shows the calculated sinusoidal short-circuit current $I_{sc}$ with an amplitude of ~35 and ~17 nA at a moderate sliding speed of 0.5 and 1 m/s, respectively. A small phase difference in the source voltage can be observed (see details in the Supplementary Information). Note that the weak interlayer vdW interaction enables the ultralow friction and even the superlubricity in 2D materials. In the experiments, an interlayer sliding speed of 25-294 m/s has been achieved[46,47]. Fig. 4c summarizes the $V_{oc}$ and $I_{sc}$ outputs as a function of the sliding speed over a wide range, from 0.001 to 10 m/s. $V_{oc}$ is independent of the sliding speed, while $I_{sc}$ depends linearly on the sliding speed. The electrical output of the tribo-FE-based NG can be tuned in a wide range by changing the sliding speed.

To quantitatively estimate the power output of the tribo-FE based NG, the NG device is regarded as a simple resistor–capacitor (RC) circuit, as was done in previous works[28,44]. It is necessary to investigate the voltage and current outputs as a function of the load resistance, as shown in Fig. 4d. Taking the sliding speed of 1 m/s as an example, the output current is unchanged for a load resistance of up to ~10 MΩ, and then decreases with increasing the load. On the other hand, the output voltage remains ~0 V initially and starts to increase at the same point. As such, the maximum delivered power of 107 nW is achieved at an intermediate load of ~350 MΩ. Regarding other sliding speeds, the RC circuit model predicts a similar behavior for the voltage and current outputs. Fig. 4e summarizes the power output as a function of the load resistance at different sliding speeds. For sliding speeds of 1, 5, and 10 m/s, the power outputs are 107, 533, and 1057 nW, respectively. Furthermore, the derived optimal load resistances are about 350, 70, and 35 MΩ, respectively. The optimized maximum power output is found to be linearly related with the sliding speed, as shown in Fig. 4f. The difference in electrical output between bilayer SnSe and monolayer MoS₂ can be found in Supplementary Table 3. For monolayer MoS₂, a mechanical deformation of 0.5 GHz can generate an alternating current with the same frequency and a power output of 0.016 nW[45], while a moderate mechanical sliding speed of 1 m/s can generate a current of 2.29 GHz and a power of 106.6 nW for bilayer SnSe.

It is worth to make a further comparison between the proposed tribo-FE-based NG and the well-known PENGs/TENGs. In addition to the high polarization change and charging rate, for the tribo-

FE based NGs, a moderate mechanical sliding speed can generate high-frequency alternating electrical output without the need for any high-frequency input signal. For example, a 1 m/s sliding speed theoretically corresponds to a current of 35 nA and a frequency of ~2.29 GHz, which is nearly impossible for PENGs. Note that, in TENGs, increasing the number of grating units of dielectrics is a popular method to generate multiple electric outputs in one directional motion cycle[44]. However, these grating units requires sophisticated microfabrication techniques, and their sizes are of several micrometers. In the proposed tribo-FE-based NG, every unit cell (length of several angstroms) can be viewed as a grating unit. Moreover, the fragile grating units in TENGs may suffer mechanical failure (e.g., wear) under high-speed sliding and consequently a short lifetime[44]. The proposed tribo-FE-based NG clearly does not have these issues. The ultralow friction between vdW layers can enhance its lifetime.

## Conclusions

In summary, a tribo-FE phenomenon was discovered in bilayer MX (M = Ge, Sn; X = S, Se) using DFT calculations. Among the four possible types of high-symmetry stacking order for bilayer MX, the AA and AC stacking orders prefer an FE state with a parallel arrangement of the polarization, while the AB and AD stacking orders favor an AFE state with an anti-parallel arrangement of the polarization. Changing the stacking order through tribological interlayer sliding results in a reversible and hysteretic phase transition between the FE and AFE states. An in-depth analysis revealed that the lattice distortion caused by the interlayer vdW force plays a dominant role in the phase transition, which is surprising. Inspired by the intrinsically high magnitude of the electric polarization change and ultrahigh changing rate, we propose a tribo-FE bilayer MX-based energy harvesting NG. The performance predicted using a well-established theoretical model is superior compared with those of the widely studied PENG and TENG devices. This tribo-FE bilayer MX-based device is a promising candidate for future NGs.

## Methods

The DFT calculations in this work were performed using the Vienna ab initio simulation package (VASP). The projector-augmented wave (PAW) potentials with the generalized gradient approximation in the Perdew–Burke–Ernzerhof (GGA–PBE) formulation were used with a cutoff energy of 600 eV. The Brillouin zone integration for structure relaxation was obtained using a 25×25×1 k-point grid. For the total energy calculation, the Brillouin zone was sampled with 45×45×1 k-points. A 30-Å-thick vacuum region was introduced to avoid interaction between the bilayers. The DFT-D2 method of Grimme was applied for the vdW corrections, and the zero damping DFT-D3 method of Grimme[48] was also applied to further check the stability of bilayer MX (Supplementary Figure 11). To verify the little impact of a substrate on the tribo-FE phenomenon, a substrate clamped model was also considered, as shown in Supplementary Figure 12. The convergence criteria for electronic and ionic relaxations were $10^{-6}$ eV and $10^{-3}$ eV/Å, respectively. The electric polarization was computed based on the Berry-phase theory of polarization[49]. To distinguish a stacking order from others, the relative position of a pair of atoms from the top and bottom layers was employed. The selective dynamics tag of VASP was used to fix the position of this pair of atoms. Further calculation details can be found in the Supplementary Information.

**Acknowledgements**

The authors gratefully acknowledge the support of NSFC (Grant Nos.11974269, 51728203), and the support by 111 project 2.0 (Grant No. BP2018008). J. D. also thanks the support of National Key R&D Program of China (Grant No. 2018YFB1900104). J.Z.L. acknowledges the support from ARC discovery projects (DP180101744) and HPC from National Computational Infrastructure from Australia. This work is also supported by State Key Laboratory for Mechanical Behavior of Materials. The authors also thank Dr. X. D. Zhang and Mr. F. Yang at Network Information Center of Xi'an Jiaotong University for support of HPC platform.

**References**

1. Guan, Z. *et al.* Recent progress in two-dimensional ferroelectric materials. *Adv. Electron. Mater.* **6**, 1900818 (2020).
2. Chang, K. *et al.* Discovery of robust in-plane ferroelectricity in atomic-thick SnTe. *Science* **353**, 274–278 (2016).
3. Belianinov, A. *et al.* $CuInP_2S_6$ Room temperature layered ferroelectric. *Nano Lett.* **15**, 3808–3814 (2015).
4. Zhou, Y. *et al.* Out-of-plane piezoelectricity and ferroelectricity in layered α-$In_2Se_3$ nanoflakes. *Nano Lett.* **17**, 5508–5513 (2017).
5. Fei, Z. *et al.* Ferroelectric switching of a two-dimensional metal. *Nature* **560**, 336–339 (2018).
6. Yuan, S. *et al.* Room-temperature ferroelectricity in $MoTe_2$ down to the atomic monolayer limit. *Nat. Commun.* **10**, 1775 (2019).
7. Shirodkar, S. N. & Waghmare, U. V. Emergence of ferroelectricity at a metal-semiconductor transition in a 1T monolayer of $MoS_2$. *Phys. Rev. Lett.* **112**, (2014).
8. Ding, W. *et al.* Prediction of intrinsic two-dimensional ferroelectrics in $In_2Se_3$ and other $III_2$-$VI_3$ van der Waals materials. *Nat. Commun.* **8**, 14956 (2017).

9. Xiao, C. *et al.* Elemental ferroelectricity and antiferroelectricity in group-V monolayer. *Adv. Funct. Mater.* **28**, 1707383 (2018).
10. Wang, H. & Qian, X. Two-dimensional multiferroics in monolayer group IV monochalcogenides. *2D Mater.* **4**, 015042 (2017).
11. Chen, W. *et al.* Direct observation of van der Waals stacking–dependent interlayer magnetism. *Science* **366**, 983–987 (2019).
12. Wang, Y. *et al.* Stacking-dependent optical conductivity of bilayer graphene. *ACS Nano* **4**, 4074–4080 (2010).
13. Kim, C.-J. *et al.* Stacking order dependent second harmonic generation and topological defects in *h*-BN bilayers. *Nano Lett.* **13**, 5660–5665 (2013).
14. Dai, J. & Zeng, X. C. Bilayer phosphorene: effect of stacking order on bandgap and its potential applications in thin-film solar cells. *J. Phys. Chem. Lett.* **5**, 1289–1293 (2014).
15. Liu, X., Pyatakov, A. P. & Ren, W. Magnetoelectric coupling in multiferroic bilayer $VS_2$. *Phys. Rev. Lett.* **125**, 247601 (2020).
16. Li, L. & Wu, M. Binary compound bilayer and multilayer with vertical polarizations: two-dimensional ferroelectrics, multiferroics, and nanogenerators. *ACS Nano* **11**, 6382–6388 (2017).
17. Xiao, J. *et al.* Berry curvature memory through electrically driven stacking transitions. *Nat. Phys.* **16**, 1028–1034 (2020).
18. Liang, Y., Shen, S., Huang, B., Dai, Y. & Ma, Y. Intercorrelated ferroelectrics in 2D van der Waals materials. *Mater. Horiz.* 10.1039.D1MH00446H (2021) doi:10.1039/D1MH00446H.
19. Yang, Q., Wu, M. & Li, J. Origin of two-dimensional vertical ferroelectricity in $WTe_2$ bilayer and multilayer. *J. Phys. Chem. Lett.* **9**, 7160–7164 (2018).
20. Liu, C., Wan, W., Ma, J., Guo, W. & Yao, Y. Robust ferroelectricity in two-dimensional SbN and BiP. *Nanoscale* **10**, 7984–7990 (2018).
21. Wu, M. & Zeng, X. C. Bismuth oxychalcogenides: a new class of ferroelectric/ferroelastic materials with ultra high mobility. *Nano Lett.* **17**, 6309–6314 (2017).
22. Fei, R., Kang, W. & Yang, L. Ferroelectricity and phase transitions in monolayer group-IV monochalcogenides. *Phys. Rev. Lett.* **117**, (2016).
23. Hu, T. & Kan, E. Progress and prospects in low-dimensional multiferroic materials. *WIREs Comput. Mol. Sci.* **9**, e1409 (2019).
24. Wu, M. & Zeng, X. C. Intrinsic ferroelasticity and/or multiferroicity in two-dimensional phosphorene and phosphorene analogues. *Nano Lett.* **16**, 3236–3241 (2016).
25. Bao, Y. *et al.* Gate-tunable in-plane ferroelectricity in few-layer SnS. *Nano Lett.* **19**, 5109–5117 (2019).
26. Chang, K. *et al.* Microscopic manipulation of ferroelectric domains in SnSe monolayers at room temperature. *Nano Lett.* **20**, 6590–6597 (2020).
27. Higashitarumizu, N. *et al.* Purely in-plane ferroelectricity in monolayer SnS at room temperature. *Nat. Commun.* **11**, (2020).
28. Wu, W. *et al.* Piezoelectricity of single-atomic-layer $MoS_2$ for energy conversion and piezotronics. *Nature* **514**, 470–474 (2014).
29. Liu, Z. *et al.* Observation of microscale superlubricity in graphite. *Phys. Rev. Lett.* **108**, 205503 (2012).
30. Hod, O., Meyer, E., Zheng, Q. & Urbakh, M. Structural superlubricity and ultralow friction across the length scales. *Nature* **563**, 485–492 (2018).
31. Zhang, S., Ma, T., Erdemir, A. & Li, Q. Tribology of two-dimensional materials: from mechanisms to modulating strategies. *Mater. Today* **26**, 67–86 (2019).
32. Han, E. *et al.* Ultrasoft slip-mediated bending in few-layer graphene. *Nat. Mater.* **19**, 305–309 (2020).
33. Berman, D., Erdemir, A. & Sumant, A. V. Approaches for achieving superlubricity in two-dimensional materials. *ACS Nano* **12**, 2122–2137 (2018).

34. Zhang, J.-J., Guan, J., Dong, S. & Yakobson, B. I. Room-temperature ferroelectricity in group-IV metal chalcogenide nanowires. *J. Am. Chem. Soc.* **141**, 15040–15045 (2019).
35. Mehboudi, M. *et al.* Structural phase transition and material properties of few-layer monochalcogenides. *Phys. Rev. Lett.* **117**, (2016).
36. Chang, K. *et al.* Enhanced spontaneous polarization in ultrathin SnTe films with layered antipolar structure. *Adv. Mater.* **31**, 1804428 (2019).
37. Kaloni, T. P. *et al.* From an atomic layer to the bulk: low-temperature atomistic structure and ferroelectric and electronic properties of SnTe films. *Phys. Rev. B* **99**, 134108 (2019).
38. Choi, K. J. Enhancement of ferroelectricity in strained BaTiO$_3$ thin films. *Science* **306**, 1005–1009 (2004).
39. Wu, X., Rabe, K. M. & Vanderbilt, D. Interfacial enhancement of ferroelectricity in CaTiO$_3$/BaTiO$_3$ superlattices. *Phys. Rev. B* **83**, (2011).
40. Reimers, J. R., Tawfik, S. A. & Ford, M. J. van der Waals forces control ferroelectric–antiferroelectric ordering in CuInP$_2$S$_6$ and CuBiP$_2$Se$_6$ laminar materials. *Chem. Sci.* **9**, 7620–7627 (2018).
41. Tawfik, S. A., Reimers, J. R., Stampfl, C. & Ford, M. J. van der Waals forces control the internal chemical structure of monolayers within the lamellar materials CuInP$_2$S$_6$ and CuBiP$_2$Se$_6$. *J. Phys. Chem. C* **122**, 22675–22687 (2018).
42. Wang, Z. L. On Maxwell's displacement current for energy and sensors: the origin of nanogenerators. *Mater. Today* **20**, 74–82 (2017).
43. Lee, J.-H. *et al.* Reliable piezoelectricity in bilayer WSe$_2$ for piezoelectric nanogenerators. *Adv. Mater.* **29**, 1606667 (2017).
44. Niu, S. & Wang, Z. L. Theoretical systems of triboelectric nanogenerators. *Nano Energy* **14**, 161–192 (2015).
45. Zhou, Y. *et al.* Theoretical study on two-dimensional MoS$_2$ piezoelectric nanogenerators. *Nano Res.* **9**, 800–807 (2016).
46. Yang, J. *et al.* Observation of high-speed microscale superlubricity in graphite. *Phys. Rev. Lett.* **110**, 255504 (2013).
47. Peng, D. *et al.* Load-induced dynamical transitions at graphene interfaces. *Proc. Natl. Acad. Sci.* **117**, 12618–12623 (2020).
48. Grimme, S., Antony, J., Ehrlich, S. & Krieg, H. A consistent and accurate *ab initio* parametrization of density functional dispersion correction (DFT-D) for the 94 elements H-Pu. *J. Chem. Phys.* **132**, 154104 (2010).
49. King-Smith, R. D. & Vanderbilt, D. Theory of polarization of crystalline solids. *Phys. Rev. B* **47**, 1651–1654 (1993).

# Figure 1

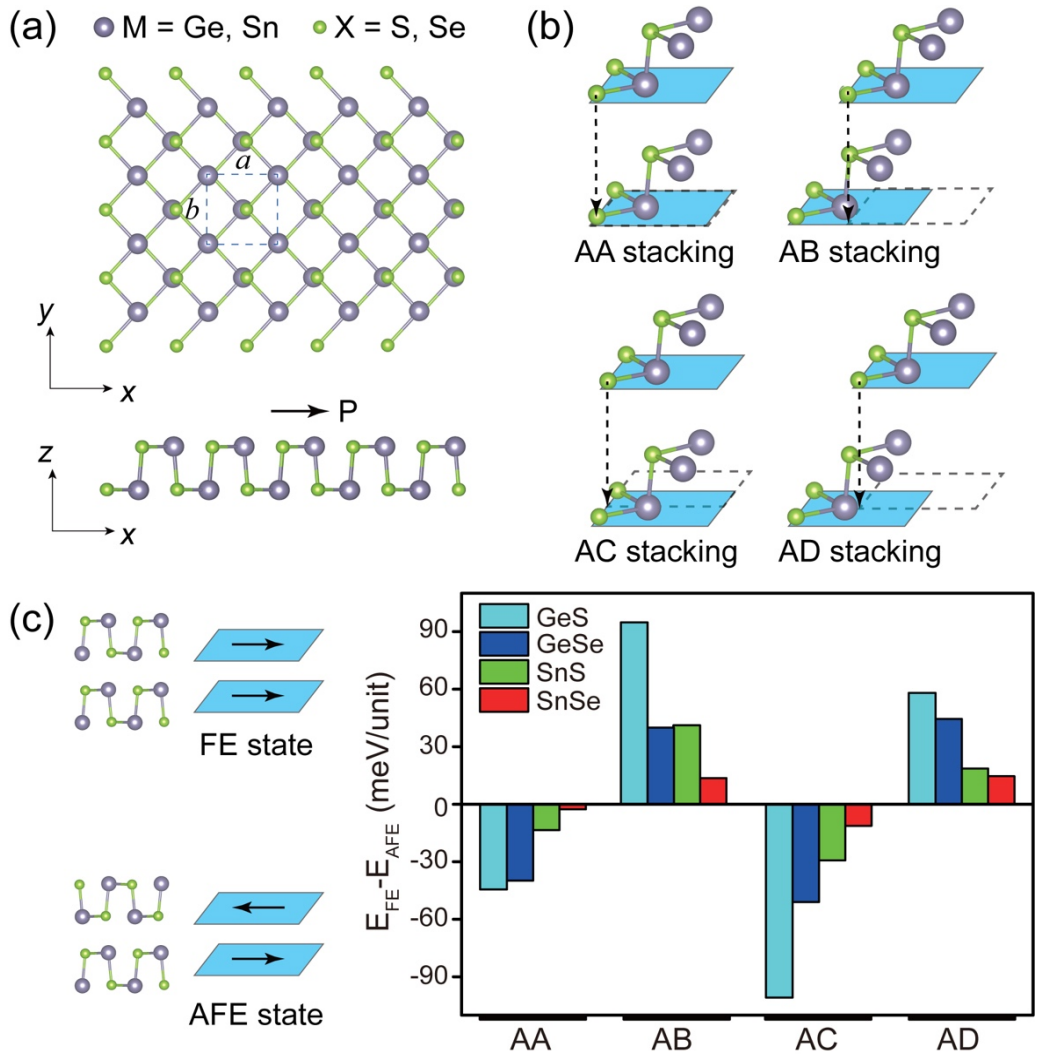

**Fig. 1 Bilayer MX (M = Ge, Sn; X = S, Se) with four different stacking orders exhibiting a ferroelectric (FE) or antiferroelectric (AFE) state. a** Monolayer MX crystal structure with a built-in spontaneous FE polarization along the armchair (*x*-) direction. **b** Four high-symmetry stacking orders of bilayer MX. Placing the top layer onto the (0, 0), (0.5, 0), (0, 0.5), and (0.5, 0.5) positions in the *x-y* plane with respect to the bottom layer leads to the AA, AB, AC, and AD stacking orders, respectively. **c** The FE polarization of the top and bottom layers can be parallel or anti-parallel, leading to the FE or AFE state, respectively. The total energy differences between the FE and AFE states for bilayer MX are illustrated for the four different stacking orders. For all investigated MX materials, the AA and AC stacking orders prefer the FE state, whereas the AB and AD stacking orders prefer the AFE state.

# Figure 2

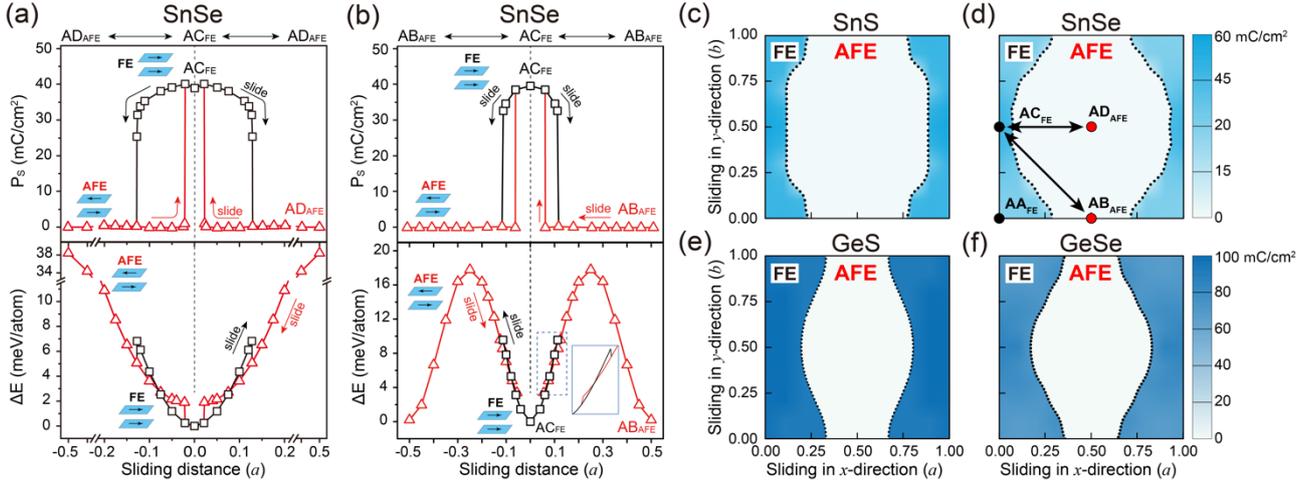

**Fig. 2 FE/AFE phase transition upon tribological mechanical interlayer sliding denoted as tribo-ferroelectricity. a** FE polarization $P_s$ and total energy difference as functions of the sliding distance (along the $x$-direction) of the top layer from the AC to the AD stacking order for bilayer SnSe. The sudden drop in the increase of $P_s$ at the critical sliding distance of $0.128a$ or $0.021a$ indicates the spontaneous FE-to-AFE or AFE-to-FE phase transition, respectively. A hysteresis loop can be observed, which is analogous to that of the traditional FE materials under external electric field stimulus. Note that the AC stacking has a stable FE state, while the AD stacking has a stable AFE state (Fig. 1c). At the two critical transition points, bilayer SnSe suddenly releases energy and transforms into the more stable phase. In fact, the energy crossing point is between the two critical points. The apparently delayed phase transition can be attributed to the energy barrier, which will be discussed later. **b** $P_s$ and total energy difference as functions of the sliding distance (along the diagonal direction in the $x$–$y$ plane) from the AC stacking to the AB stacking. Similar phase transitions and hysteresis loop can be observed. A double-well-like curve due to the metastable AB stacking is shown (different from the unstable AD stacking). **c–f** Contour plots of $P_s$ as a function of the mechanical sliding distance (leading to different stacking orders) for SnS, SnSe, GeS, and GeSe, respectively. The color bar denotes the $P_s$ magnitude of bilayer MX. The black dotted lines represent the phase boundary between the FE and AFE states. The tribological mechanical sliding of the top layer across the boundary leads to the reversible FE/AFE phase transition, i.e., reversible switching of the FE polarization. This new phenomenon is here referred to as tribo-FE.

# Figure 3

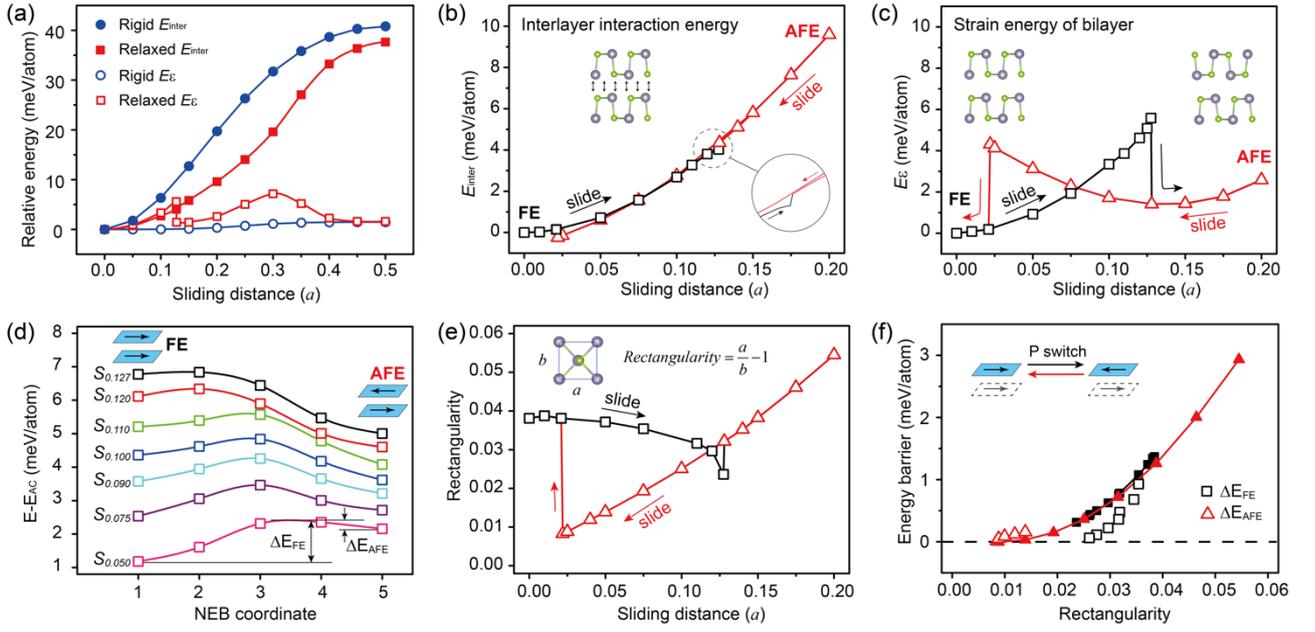

**Fig. 3 Physical origin of the spontaneous tribo-FE phenomenon. a** Comparison of $E_{inter}/E_\varepsilon$ of the relaxed and rigid bilayer SnSe upon mechanical sliding of the top layer. The significant interlayer energy $E_{inter}$ reduction dominates over the lattice strain energy increase in the relaxed SnSe bilayer, thus driving the lattice distortion. **b** $E_{inter}$ energy and **c** $E_\varepsilon$ energy of the relaxed bilayer SnSe upon mechanical sliding from $AC_{FE}$ to $AD_{AFE}$. The small $E_{inter}$ changes between the FE and AFE states at the phase transition points are opposite to the total energy change in Fig. 2a. As the sliding distance from $AC_{FE}$ increases, $E_\varepsilon$ increases and exhibits a sudden drop at the FE-to-AFE transition. In the reverse sliding process, $E_\varepsilon$ first decreases and then increases before suddenly dropping at the AFE-to-FE transition. The change of $E_\varepsilon$ is consistent with that of $E_{tot}$ with a similar magnitude, indicating that the strain energy change is the origin for the observed spontaneous tribo-FE effect in bilayer SnSe. A similar conclusion can also be drawn for bilayer SnS. **d** Calculated energy barrier between the FE and AFE states of bilayer SnSe for different sliding positions along the pathway from $AC_{FE}$ to $AD_{AFE}$. $S_{0.050}$ denotes sliding at the position $x = 0.05a$. The energy barrier gradually vanishes. At a sliding distance of $0.09a$ (i.e., $S_{0.09}$), the FE state has a higher total energy than the AFE state, but an energy barrier still exists. This is the reason for the delayed FE-to-AFE phase transition at $x = 0.128a$ which results in the observed hysteresis in Fig. 2a. **e** Variation of the unit cell rectangularity of bilayer SnSe during the sliding process from $AC_{FE}$ to $AD_{AFE}$. The spontaneous phase transition takes place at minimum values of the rectangularity, suggesting the correlation. **f** The phase transition energy barrier depends on the unit cell rectangularity for both bilayer SnSe and monolayer SnSe. Upon reducing the rectangularity, the energy barrier decreases and eventually disappears. The agreement between bilayer and monolayer SnSe indicates that the sliding-induced lattice distortion plays a primary role in the energy barrier.

# Figure 4

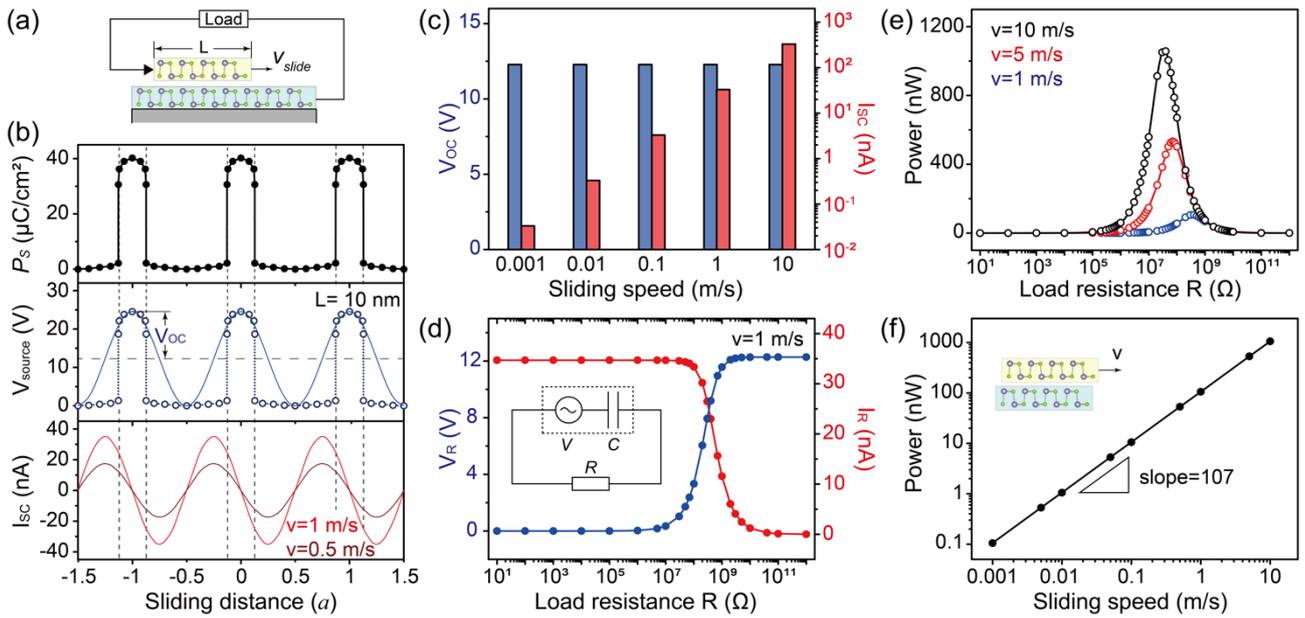

**Fig. 4 NG using the tribo-FE phenomenon – theoretical model prediction. a** Schematic illustration of a tribo-FE-based NG in connection with an external load resistor. The theoretical models were adopted from Ref. 44 (see the main text and the SI for details). **b** Variation of polarization, source voltage ($V_{source}$), and short-circuit current ($I_{sc}$) of a 10 × 10 nm$^2$ tribo-FE-based NG under a 1 m/s sliding speed. Three sliding periods are shown. The considerably large values of $V_{source}$ and $I_{sc}$ benefit from the large polarization change and large polarization rate during the interlayer sliding process (see the main text for details). The voltage exhibits a linear relation with the polarization change and is simplified as a sinusoidal alternating voltage in subsequent modelling and discussion. **c** Dependence of $V_{oc}$ and $I_{sc}$ on the interlayer sliding speed. The voltage is independent of the sliding speed because of the constant polarization change (see the main text). $I_{sc}$ is linearly related with the sliding speed because of its linear relation with the rate of polarization change. **d** Dependence of the voltage and current output under a 1 m/s sliding speed as a function of the load resistance. The inset is the equivalent circuit of the tribo-FE-based NG. **e** Power output as a function of the load resistance at different sliding speeds. **f** Maximum power at different sliding speeds obtained from the theoretical prediction. The wide range of sliding speeds attainable in 2D tribology endows this tribo-FE-based NG with a diverse and superior output power.